\documentclass[letterpaper,twocolumn,aps,prl,groupedaddress,superscriptaddress]{revtex4}
\usepackage[utf8]{inputenc}
\usepackage{amstext}
\usepackage{graphicx}

\makeatletter
\@ifundefined{textcolor}{}
{%
 \definecolor{BLACK}{gray}{0}
 \definecolor{WHITE}{gray}{1}
 \definecolor{RED}{rgb}{1,0,0}
 \definecolor{GREEN}{rgb}{0,1,0}
 \definecolor{BLUE}{rgb}{0,0,1}
 \definecolor{CYAN}{cmyk}{1,0,0,0}
 \definecolor{MAGENTA}{cmyk}{0,1,0,0}
 \definecolor{YELLOW}{cmyk}{0,0,1,0}
}

\@ifundefined{definecolor}{\usepackage{color}}{}
\@ifundefined{definecolor}{\usepackage{color}}{}
\@ifundefined{definecolor}{\usepackage{color}}{}

\usepackage{dcolumn}
\usepackage{bm}
\@ifundefined{definecolor}{\usepackage{color}}{} 
\usepackage{xcolor}

\makeatother

\begin{document}

\title{Quasi-2D magnetism and origin of the Dirac semimetallic behavior in nonstoichiometric Sr$_{1-y}$Mn$_{1-z}$Sb$_{2}$ (y, z~$<$0.1)}

\author{Qiang Zhang}
 \email{zhangq6@ornl.gov}
\affiliation{Department of Physics and Astronomy, Louisiana State University, Baton Rouge, Louisiana 70803, USA}
\affiliation{Neutron Scattering Division, Oak Ridge National Laboratory, Oak Ridge, Tennessee 37831, USA}

\author{Satoshi Okamoto}

\affiliation{Materials Science and Technology Division, Oak Ridge National Laboratory, Oak Ridge, Tennessee 37831, USA}

\author{Matthew B. Stone}

\affiliation{Neutron Scattering Division, Oak Ridge National Laboratory, Oak Ridge, Tennessee 37831, USA}

\author{Jinyu Liu}

\affiliation{Department of Physics and Engineering Physics, Tulane University, New Orleans, Louisiana 70118, United States, USA}

\author{Yanglin Zhu}

\affiliation{Department of Physics and Engineering Physics, Tulane University, New Orleans, Louisiana 70118, United States, USA}

\author{John DiTusa}
\affiliation{Department of Physics and Astronomy, Louisiana State University, Baton Rouge, Louisiana 70803, USA}

  \author{Zhiqiang Mao} 

\affiliation{Department of Physics and Engineering Physics, Tulane University, New Orleans, Louisiana 70118, United States, USA}
\affiliation{Department of Physics, Pennsylvania State University, University Park, PA 16802}

\author{David Alan Tennant}
 \email{tennantda@ornl.gov}
\affiliation{Materials Science and Technology Division, Oak Ridge National Laboratory, Oak Ridge, Tennessee 37831, USA}
\affiliation{Shull Wollan Center, Oak Ridge National Laboratory, Oak Ridge, Tennessee 37831, USA}

\date{\today}
\begin{abstract}

 Nonstoichiometric Sr$_{1-y}$Mn$_{1-z}$Sb$_{2}$ (y, z~$<$0.1) is known to exhibit a coexistence of 
 magnetic order
 and the nontrivial semimetallic behavior related to Dirac or Weyl fermions.
Here, we report inelastic neutron scattering analyses of the spin dynamics and density functional
theory studies on the electronic properties of Sr$_{1-y}$Mn$_{1-z}$Sb$_{2}$. We observe a 
relatively large spin excitation gap $\approx$ 8.5 meV at 5 K, and the interlayer magnetic 
exchange constant only 2.8 \% of the dominant intralayer magnetic interaction, providing 
evidence that Sr$_{1-y}$Mn$_{1-z}$Sb$_{2}$ exhibits a quasi-2D magnetism. Using density functional
theory, we find a strong influence of magnetic orders on the electronic band structure and
the Dirac dispersions near the Fermi level along the Y-S direction in the presence of a ferromagnetic ordering.
Our study unveils novel interplay between the magnetic order, magnetic transition, and 
electronic property in Sr$_{1-y}$Mn$_{1-z}$Sb$_{2}$, and opens new pathways to control the relativistic band structure through magnetism in ternary compounds.

    \end{abstract}


 \maketitle
Topological semimetals \cite {Novoselov2005,Armitage2018} are a newly emerged frontier in condensed matter physics and have stimulated tremendous research interest because they give access to new quantum phenomena and 
are very attractive for both fundamental research and technological application. Particular 
attention has focused on Weyl or Dirac semimetals that exibit a coexistence with magnetism as a promising route to 
modify and control Weyl/Dirac fermions, electronic transport properties and band topology, among which SrMnSb$_{2}$ has attracted much interest recently.
Liu \textit{et al.} \cite{liu2017} reported a nontrivial semimetallic
behavior related to Dirac or Weyl fermions including nearly massless quasiparticles with a $\pi$ Berry
phase coupled to ferromagnetism in nonstoichiometric Sr$_{1-y}$Mn$_{1-z}$Sb$_{2}$. For the
magnetic behavior, it displays ferromagnetic (FM) order below $T_{C} \sim$ 565 K,
followed by a transition to canted antiferromagnetic (AFM) order with
a net FM component below $T_{FM-AFM} \approx$ 304 K \cite{liu2017}. The nontrivial topological semimetal behavior was further supported by optical conductivity and ultrafast optical pump-probe measurements\cite{Weber2017}. 
   Nevertheless, Ramankutty \textit{et al.} \cite{Ramankutty2018} reported
   zero Berry phase indicative of trivial topology in nearly stoichiometric SrMnSb$_{2}$. Previous density functional theory (DFT) calculations\cite{Ramankutty2018,Farhan2014,You2018} showed that the lattice distortion in the orthorhombic structure
   prevents the formation of the Dirac
   points near the Fermi level by opening a gap. Both results cannot account for the observed topological 
   semimetallic behavior reported in \textit{Ref}. \cite{liu2017}. It is therefore challenging 
   to explore whether there are indeed Dirac/Weyl points in proximity to the Fermi
level and what drives their formation in Sr$_{1-y}$Mn$_{1-z}$Sb$_{2}$. Furthermore, SrMnSb$_{2}$ offers a wonderful opportunity to address an important question whether there is a close correlation between magnetic order and band topology in 3D Dirac compounds.

   In addition to SrMnSb$_{2}$, other Alkaline earth ternary AMnC$_{2}$
 ``112''  compounds (A =Sr, Ca, Ba, C= Bi or Sb) \cite{Wang2011,Ray2017,Rahn2017,Liu2016,Huang2017}
 were reported to be Dirac semimetal candidates  
 with the coexistence of AFM order. An interplay between magnetic order and electronic transport properties was found 
 in CaMnBi$_{2}$ \cite{Guo2014} due to coupling of the interlayer ferromagnetic component to the planar Bi
 electrons. 
  The Dirac carriers in Bi layers were reported to enhance the interlayer exchange coupling $J_{\perp}$ significantly between magnetic layers 
 in AMnBi$_{2}$ (A=Ca, Sr) by the Raman spectroscopy \cite{AZhang2015}. However, Rahn\textit{et al.} \cite {Rahn2017} argued that the neglect of single-ion anisotropy \textit{D} in the Raman 
 analysis may significantly exaggerate the obtained interlayer 
 magnetic coupling since $D$ and $J_{\perp}$ are correlated. All of these facts emphasize
 the importance of an accurate determination of the interlayer magnetic coupling and the magnetic dimensionality in
 ``112'' compounds.

   Here, by a combination of inelastic neutron scattering and linear spin wave theory, we report the magnetic excitation spectra and determination of the accurate 
   magnetic exchange couplings and the single-ion anisotropy, which evidences a quasi-2D magnetism in  Sr$_{1-y}$Mn$_{1-z}$Sb$_{2}$. More interestingly, we found a strong 
   coupling between various magnetic orders 
   and the relativistic band structures near the Fermi level by density functional theory (DFT), which reveals that FM order with the moment along \textit{b} axis induces
   the Dirac points near the Fermi
   level in the band structure whereas various AFM 
   orders have a disfavoring effect on it in Sr$_{1-y}$Mn$_{1-z}$Sb$_{2}$.

    Sr$_{1-y}$Mn$_{1-z}$Sb$_{2}$ crystals were grown using a flux technique \cite{liu2017}.
Several single crystals with a total mass of approximately
600 mg were co-aligned at the (0 \textit{K} \textit{L}) horizontal scattering plane within
$\sim3$ degrees mosaicity.  Inelastic neutron measurements
 were performed using the Spallation Neutron Source's SEQUOIA spectrometer with its high-flux mode at Oak Ridge National Laboratory.  The data were collected at 5 K and 350 K using a few different incident energies of 35, 70, 100, 160, 200 meV.
The wave vectors \textbf{Q} reported here are defined in reciprocal lattice unit (rlu).
The constant-energy ($E$) cuts were fitted using a Lorentz function to obtain both the spin wave 
dispersion and intensity. The fits to spin wave dispersion and intensity using the SpinW package \cite{Toth2015} yield the magnetic exchange 
constants and single-ion anisotropy. DFT was performed using the generalized gradient approximation and projector augmented wave approach \cite{Blochl1994} as implemented in the Vienna {\it ab} initio simulation package (VASP) \cite{Kresse1996,Kresse1999}.

\begin{figure}
\centering \includegraphics[width=1\linewidth]{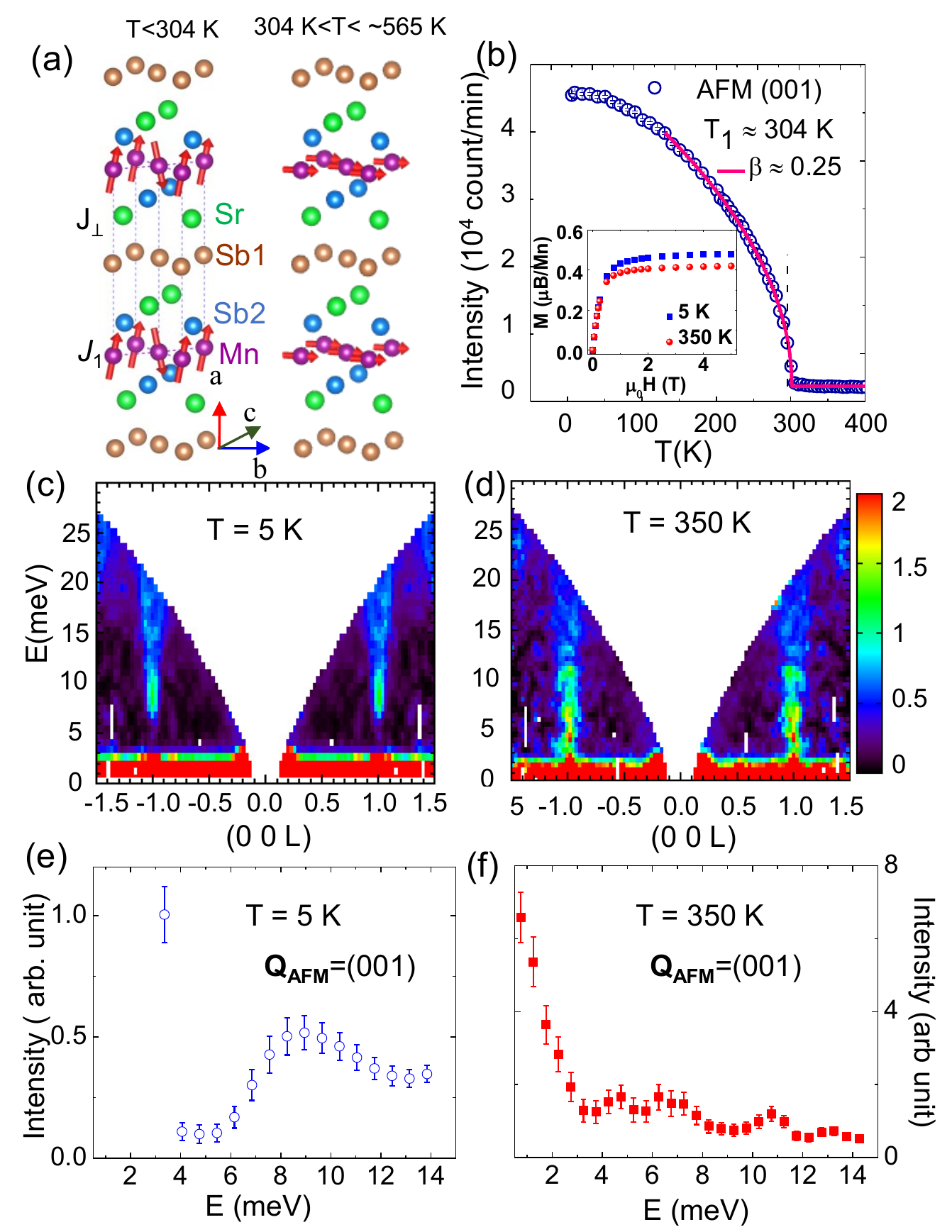} \caption{ (color online) (a). Magnetic structures of Sr$_{1-y}$Mn$_{1-z}$Sb$_{2}$: canted AFM order in $T<$304 K and FM order with moment along $b$ axis in 304 K $<T<\sim$ 565 K.
(b). Temperature dependence of the peak intensity for the pure AFM Bragg peak (001) taken from \textit{Ref}. \cite{liu2017}. The solid line is the fit to the powder law. Inset shows the field dependence of the magnetization at different temperatures. Magnetic excitations near AFM zone center (0 0 $\pm$1) at (c) 5 K and  (d) 350 K. The 
corresponding constant-Q cuts at 5 K and 350 K are shown in (e) and (f), respectively.
}

\label{fig:SpinGap} 
\end{figure}

   Figure~\ref{fig:SpinGap}(a) shows the crystal and magnetic structures in Sr$_{1-y}$Mn$_{1-z}$Sb$_{2}$. It crystallizes in the orthorhombic structure with space group \textit{Pnma} (No. 62), 
  consisting of a MnSb layer with edge-sharing MnSb(2)$_{4}$ tetrahedral and flat Sb(1) layer sandwiched between
   two staggered Sr planes. Note that all the Sb(1) square net, Sr and MnSb(2)$_{4}$ are distorted,
   different from the tetragonal structure in Bi-based ``112''  compounds. The two magnetic structures
   previously determined \cite{liu2017} for $T<T_{FM-AFM}$ and $T_{FM-AFM}<T<T_{C}$ are illustrated in Fig. \ref{fig:SpinGap} (a). Fits to the order parameter 
   of the magnetic Bragg peak (001) in Fig. \ref{fig:SpinGap} (b) to a power law of the fom $I\propto(T_{N}-T)^{2\beta}$ 
   yields a critical exponent $\beta\approx$ 0.25. Such a $\beta$ value falls in the region
   $0.1< \beta <0.25$ expected for quasi-2D systems \cite{Taroni2008} (compared to the 0.36 in the 3D Heisenberg model) and is similar to that in typical quasi-2D pnictide LnMnSbO (Ln=La or Ce) \cite{Zhang2016}. This implies that 
    Sr$_{1-y}$Mn$_{1-z}$Sb$_{2}$ may be magnetically quasi two-dimensional. The field dependence of 
    the magnetization in the inset of Fig. 1 (b) confirms a clear ferromagnetism below $T<T_{C}$ down to 5 K.

\begin{figure}
\centering \includegraphics[width=1\linewidth]{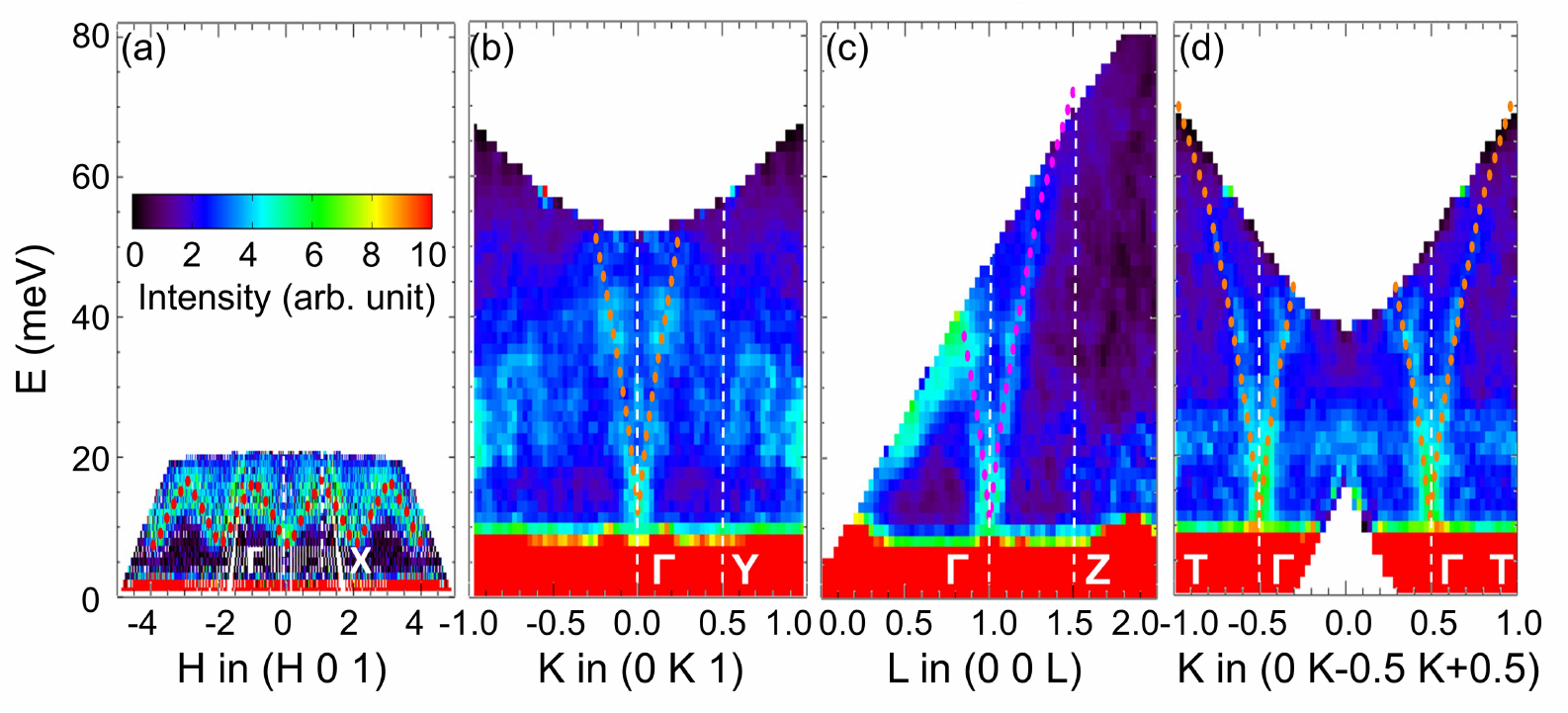} \caption{(color online) 
Magnetic excitations near AFM zone center (0 0 1) at 5 K, along high-symmetry directions: (a)
out-of-plane\textit{ H}, in-plane (b) \textit{ K}, (c) \textit{ L} and (d) diagonal [0 \textit{K} \textit{K}] directions. The dashed lines are fits to the experimental spin wave
dispersions. The vertical lines indicate the AFM zone center or boundaries, with the brillouin symbols marked in the figures.}
\label{fig:Escan} 
\end{figure}

       To investigate the spin dynamics of Sr$_{1-y}$Mn$_{1-z}$Sb$_{2}$, we performed inelastic neutron 
       scattering measurements. Figure 1 (c,e) and
       (d,f) compares the magnetic excitations near the AFM zone center (0 0 1) at 5 K 
       ($T<T_{FM-AFM}$) and 350 K ($T_{FM-AFM}<T<T_{C}$). At 5 K within the AFM ordered state, 
a clear spin gap $E_{g}\approx$ 8.5 meV is observed indicative of the existence of a single-ion
anisotropy.  The spin gap closes at 350 K when
the long-range AFM order disappears. Another difference between these two temperatures is
that, whereas the spin wave dispersion exists at 5 K, the dispersion disappears, and evolves into a spin fluctuation spectra 
near the AFM zone centers at 350 K. The magnetic excitations along
out-of-plane \textit{H}, in-plane \textit{K}, \textit{L} and diagonal [0 \textit{K} \textit{K}] directions 
at 5 K are displayed in
Fig. \ref{fig:Escan}  (a-d) (the high-symmetric brillouin symbols are illustrated in the inset of 
Fig.~\ref{fig:theory}~(a)). There is a steep dispersion along in-plane directions extending to $\approx$ 70 meV at the AFM zone boundary
Z and T points, but the dispersion along out-of-plane \textit{H} direction is much weaker, with E $\approx$ 18 meV at the zone-boundary X point. 
This indicates that the out-of-plane magnetic interaction is much weaker than the in-plane one suggesting a quasi-2D magnetism.

\begin{figure}
\centering \includegraphics[width=0.9\linewidth]{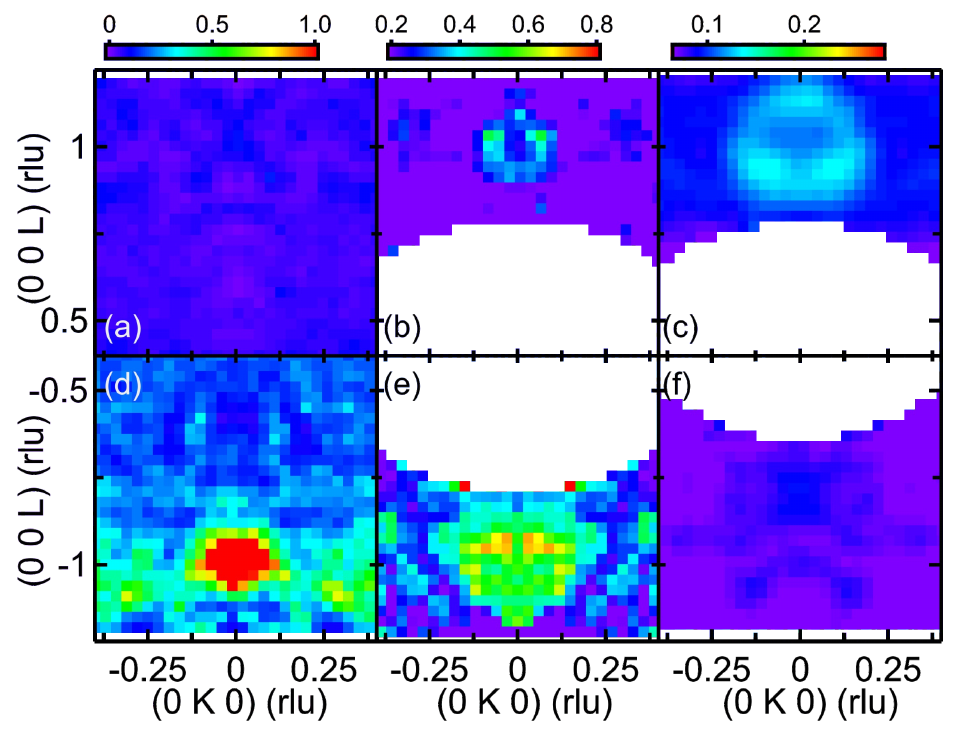} \caption{(color online) Constant-E slices in (0 \textit{K L}) scattering plane around E of  (a) 5 meV, (b) 17 meV
and (c) 30 meV at 5 K. Corresponding constant-E slices at 350 K are shown in (d-f).}
\label{fig:Qscan} 
\end{figure}

\begin{figure*}
\centering \includegraphics[width=1\linewidth]{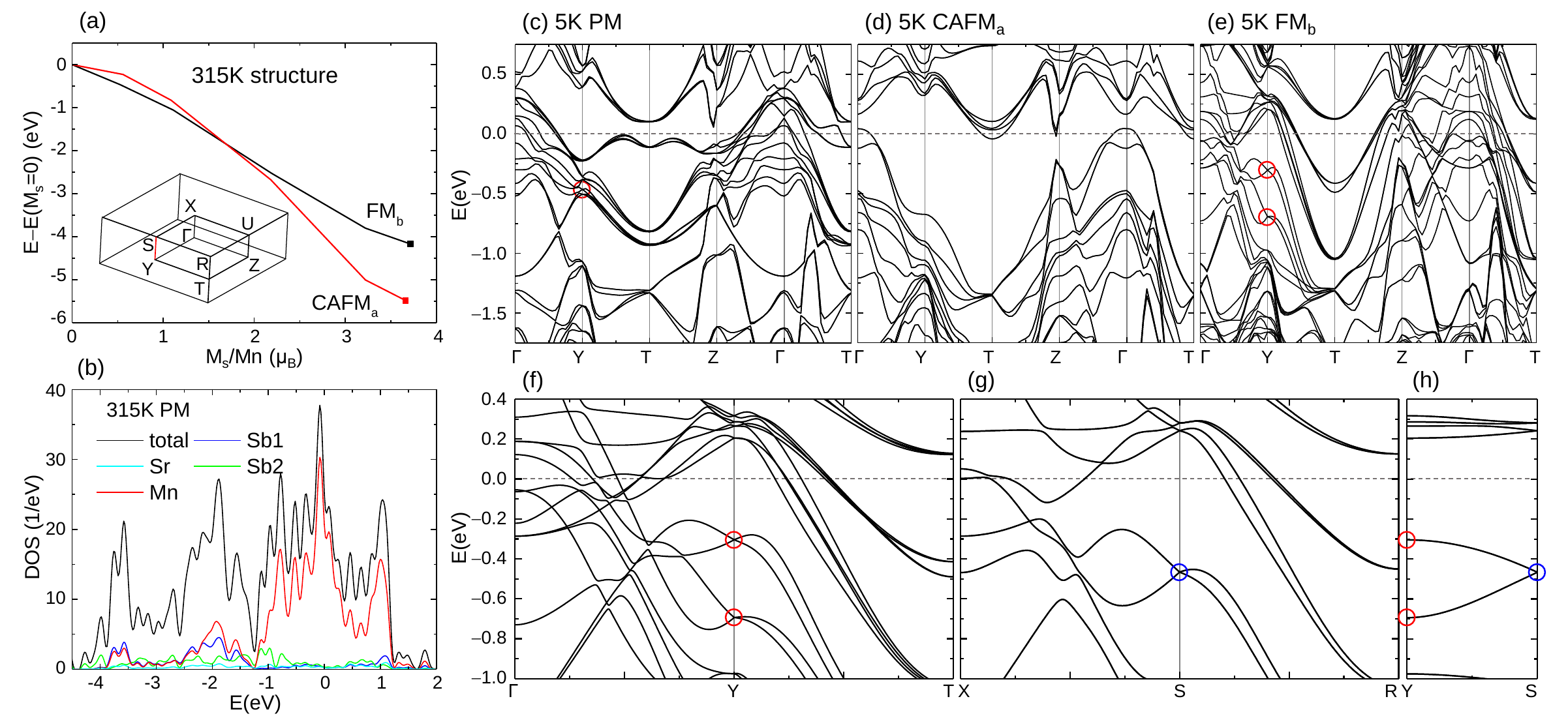} \caption{Density functional theory results. 
(a) Total E as a function of ordered moment $M_s$ on a Mn site for FM$_b$ and CAFM$_a$. Squares indicate equilibrium positions. 
(b) Density of states in the PM phase. 
Dispersion relation along high-symmetry lines for (c) PM, (d) CAFM$_a$, and (e) FM$_b$. 
Dirac dispersions are indicated by circles in (c) and (e). 
(f)-(h) are the magnified view of the Dirac dispersions in the (e) FM$_b$ state along 
different directions, 
showing the evolution of two non-degenerate Dirac cones (red circles) at the Y point merge to a twofold degenerate Dirac cone at the S point (blue circle). 
The results in (a) and (b) are obtained using the high-temperature structure at 315~K, while (c-h) using the low-temperature structure at 5~K. 
The inset of (a) shows the 1st Brillouine zone, with the Dirac line node along the Y-S direction indicated by a red line.
Energy in (b)-(h) is measured from the Fermi level. }
\label{fig:theory} 
\end{figure*}

       Figure \ref{fig:Qscan} compares constant-energy slices in the (0 \textit{K} \textit{L})
       scattering plane at energy transfers around 5, 17 and 30 meV at 5 and 350 K. At 5 K, 
     no magnetic excitations are apparent around 5 meV in Fig. \ref{fig:Qscan}(a) owing to the existence
     of higher spin gap of $E_{g}\approx$8.5 meV.  As the transferred E increases, 
     a ring of scattering emerges at AFM zone center positions
     such as (0 0 $\pm$1) and (0 $\pm$1 0). The diameter of the rings increases with increasing the E
     transfer (see Fig. \ref{fig:Qscan}(b) and (c)), indicative of dispersive spin waves. In sharp contrast,
     the magnetic excitations exhibit different features at 350 K.
     A diffuse magnetic excitations can be seen in Fig. 3(d) at around 5 meV due to the closure
     of the spin-gap. 
     As E transfer increases,
     the magnetic excitations become more diffuse and spread out without the ring-like feature, as shown
     in Fig.  \ref{fig:Qscan}  (e) and eventually evolve to be hardly visible at high E transfer region in Fig. \ref{fig:Qscan} (f). This indicates the existence of 
     the low-E AFM spin fluctuations at 350 K. It is worthwhile pointing out that we do not observe
     clear spin-wave branches associated with FM ordered phase at 350 K, 
     which is understood due to the low FM moment $\approx$ 0.41 $\mu_{B}$ (see inset of Fig. ~\ref{fig:SpinGap} (b) and small mass of the coaligned crystals ($\approx$ 600 mg).

     To analyze the observed spin waves and quantitively determine the magnetic interactions in Sr$_{1-y}$Mn$_{1-z}$Sb$_{2}$, we have performed linear spin wave calculations using the
     SpinW package \cite{Toth2015} for the following spin Hamiltonian:
\begin{equation}
H=  \sum_{i,j} S_{i}J_{ij}S_{j}+\sum_{\alpha, \beta, i} S_{i}^{\alpha} A_{i}^{\alpha,\beta} S_{i}^{\beta}
\end{equation}  where $S_{i}$  are spin vector operators, $J_{ij}$ are pair coupling between spins and $A_{i}^{\alpha,\beta}$ are 3$\times$3 anisotropy matrices. By fitting to the experimental SW dispersions and intensities, we obtain the AFM nearest-neighbor 
       (NN) $J_{1} \approx$ 9.3(3) meV, interlayer FM $J_{\perp} \approx$ -0.26(8) meV, and
 anisotropy parameter $A_{i}$=diag(0 0.12(5) 0.3(1)). Here we enumerate the key results from the SW fitting: 1). The NN $J_{1}$ is dominant and the next-nearest-neighbor (NNN) $J_{2}$ is 
 found to be negligible, i.e., $J_{2}/J_{1}<<$1/2. Together with FM $J_{\perp}$, this could account for the formation of the overall \textit{C}-type AFM 
 order \cite{Rahn2017,Zhang2016,Zhang2015}. Furthermore, the AFM NN $J_{1}$ and FM out-of-plane $J_{\perp}$ are consistent with the 
 associated AFM and FM spin arrangements (see Fig.~\ref{fig:SpinGap}~(a)), respectively. All these results 
 indicate there is no signature of strong spin frustration. 2). 
 The interlayer $J_{\perp}$, which characterizes the dispersion along the out-of-plane \textit{a} axis,
 is less than 2.8 \% of the in-plane $J_{1}$, 
 signaling the quasi-2D magnetism. The $J_{\perp}$ is considerably weaker than that proposed
 in \textit{Ref}. \cite {AZhang2015} on AMnBi$_{2}$ (A=Ca,Sr)
 (with similar interlayer Mn-Mn distances), which was claimed 
 as enhancement due to the Dirac carrier Bi layers.
 3). The emergence of the spin gap is ascribed to the uniaxial single-ion anisotropy matrices  $A_{i}^{\alpha,\beta}$.

 To gain insight into the experimental observations, we performed DFT calculations for stoichiometric SrMnSb$_2$
using the high-temperature structure at 315~K and the low-temperature one at 5~K \cite{liu2017}.
For Sr, a potential, in which semi-core $s$ and $p$ states are treated as valence states, is used  (Sr$_{sv}$),
and for Mn and Sb, standard potentials were used (Mn and Sb, respectively, in the VASP distribution). 
In most cases, we use a $2\times8\times8$ {\bf k}-point grid and an E cutoff of 500 eV with spin-orbit coupling included.
The $+U$ correction is not included because SrMnSb$_2$ is an itinerant magnetic system.

We first investigate the magnetic properties at high temperature.
Using the 315~K structure, we compute the total E of several symmetry-allowed magnetic configurations \cite{liu2017},
including \textit{C}-type AFM (CAFM), \textit{G}-type AFM, \textit{A}-type AFM, and FM with spin orientation taken along the $a$, $b$ or $c$ axis
(the spin orientation will be indicated as a subscript as for example CAFM$_a$).
It turned out  that CAFM$_a$ is most stable, while the FM$_b$ state is more stable than FM$_a$ and FM$_c$.
This discrepancy is resolved by constrained magnetization calculations.
As shown in Fig.~\ref{fig:theory}~(a), the relative E between FM$_b$ and CAFM$_a$ is reversed when the ordered Mn moment $M_s$ is suppressed
at around $M_s=1.6 \mu_{\rm B}$.
In the PM state ($Ms=0$), the density of states (DOS) has a sharp peak near the Fermi level ($E=0$) as shown in Fig.~\ref{fig:theory}~(b).
Thus, the leading instability in the high-temperature PM phase is toward FM ordering due to the Stoner instability,
followed by the first-order transition from FM$_b$ to CAFM$_a$ at lower temperature 
with much larger $Ms$ ($\approx3.6\mu_{\rm B}$ at 5 K \cite{liu2017}).
This provides a natural explanation for the sequence of magnetic transitions reported from experiments \cite{liu2017}.

Now we turn to the low-temperature electronic property using the structural information at 5 K. We computed the total energy of several magnetic configurations using the low-temperature structure at 5~K. We found it is robust 
that the in-plane (out-of-plane) exchange is AFM (FM) and the single ion anisotropy is uniaxial \cite{Satoshi2018}, which indicates that
the CAFM$_a$ state is most stable, and the FM$_b$ is metastable at 5 K. It is insightful to examine 
the electronic band structures of the PM, the CAFM$_a$ and the FM$_b$ states, as shown in Figure \ref{fig:theory}(c), (d), and (e), respectively.
Because of the non-symmorphic symmetry, Dirac cones are expected at the Y and the T points \cite{Young2015,Ramankutty2018}
(X and M points in the notation of \textit{Ref}. \cite{Ramankutty2018}).
In the PM state, the Dirac cones are observed at $E\sim -0.5$~eV at the Y point, but those at the T point are not clearly resolved in this E window.
The band structure is influenced by magnetic order.
In the CAFM$_a$ state, the Dirac cone at the T point is clearly seen at $E \sim -1.3$~eV.
This dispersion relation is consistent with the one reported in $Refs.$~\cite{Ramankutty2018,Farhan2014}
except for the current semimetallic behavior due to the difference in the exchange correlation potential or local $U$.
As discussed in \textit{Ref}.~\cite{Ramankutty2018}, the location of the Dirac cone is too far from the Fermi level to account for
the $\pi$ Berry phase reported in \textit{Ref}.~\cite{liu2017}.
Interestingly, when the metastable FM$_b$ state is considered,
the Dirac cone appearing in the PM state at the Y point splits due to the spin polarization, and one of them becomes closer to the Fermi level ($E\sim-0.3$~eV) [Fig.~\ref{fig:theory}~(e)].
As shown in Fig.~\ref{fig:theory}~(f)-(h), these Dirac cones form a line node along the Y-S direction, and two Dirac cones merge at the S point.
This is a consequence of  two units of MnSb(2)$_{4}$ and Sb$(1)$ layers, each of which supports two-dimensional Dirac cones, 
and the mixing between the two units is suppressed at $k_x=\pi$.

Once the Fermi level is tuned near the Dirac cones, the $\pi$ Berry phase should
be manifested in the quantum oscillations of magnetoresistance or magnetization. 
A natural question is how one can realize this condition in SrMnSb$_2$.
In reality, the $\pi$ Berry phase is observed only in non-stoichiometric
Sr$_{1-y}$Mn$_{1-z}$Sb$_{2}$ samples in \textit{Ref}.~\cite{liu2017}.
Thus, the current study suggests the following scenario.
The relative energy between the CAFM$_a$ and FM$_b$ in non-stoichiometric samples could be much smaller than that in stoichiometric samples.
As a result, the FM$_b$ state could remain in place more easily at low temperatures, resulting in either canted AFM order \cite{liu2017} or possible phase separation between
 collinear AFM$_a$ and FM$_b$ phases. The FM state induces the 
 splitting of Dirac cone at Y point with one of them
 being closer to the Fermi level. Moreover, the Fermi level is expected to be lowered by Sr and/or Mn 
 off-stoichiometries to locate near Dirac cones.  
In order to verify this scenario, it is necessary to control the defect density and magnetism, and isolate FM regimes from others.

   In summary, we have examined the magnetic excitations, and investigated the origin of the Dirac
   semimetallic behavior in nonstoichiometric Sr$_{1-y}$Mn$_{1-z}$Sb$_{2}$. 
   The magnetic exchange constants are determined, indicative of a quasi-2D magnetism, with no
   significant enhancement of  $J_{\perp}$ by the 
    Dirac carrier Sb layers. The constrained magnetization calculations by our DFT interpreted
 the occurrence of the successive PM-FM-AFM transition.
 We further demonstrated that while the AFM order does not favor the formation of the 
 Dirac cones near the Fermi level, the FM order/component plays a key role in inducing the Dirac points in proximity to the Fermi level to drive the system to be Dirac semimetal in Sr$_{1-y}$Mn$_{1-z}$Sb$_{2}$. Our study provides a new clue 
 to the understanding of the origin of Dirac semimetals and to seek for 
 novel Dirac semimetals by adjusting the magnetic order.

\emph{Acknowledgments}  Primary support for this study came from the U.S. Department of Energy under EPSCoR Grant No. DESC0012432,
with additional support from the Louisiana Board of Regents. A portion of this research used resources at Spallation Neutron Source,
a DOE Office of Science User Facility operated by the Oak Ridge National Laboratory. 
The research by SO and DAT was sponsored by the Laboratory Directed Research and Development Program (LDRD) of Oak Ridge National Laboratory,
managed by UT-Battelle, LLC, for the U.S. Department of Energy (Project ID 9533).


\begin{thebibliography}{10}

 \bibitem{Novoselov2005} K. S. Noselov, A. K. Geim, S. V. Morozov, D. Jiang, M. I. Katsnelson, I. V. Grigorieva, S. V. Dubosnos, and A. A. Firsov, Nature \textbf{438}, 7065 (2005).
 \bibitem{Armitage2018} N. P. Armitage, E. J. Mele, and Ashvin Vishwanath, Rev. Mod. Phys. \textbf{90}, 015001, (2018). 
 \bibitem{liu2017} J.Y. Liu, J. Hu, Q. Zhang, D. Graf , H.B. Cao, S.M.A. Radmanesh, D.J. Adams, Y.L. Zhu, G.F. Cheng, X. Liu, W. A. Phelan,
 J. Wei, M. Jaime, F. Balakirev, D. A. Tennant, J. F. DiTusa, I. Chiorescu, L. Spinu and Z.Q. Mao, Nature materials, \textbf{16}, 905 (2017).
 \bibitem{Weber2017} C. P. Weber, \textit{et al.}., J. Appl. Phys., \textbf{122}, 223102, (2017).
  \bibitem{Ramankutty2018} S. V. Ramankutty, \textit{et al.}, SciPost Phys. \textbf{4}, 010 (2018).
      \bibitem{Farhan2014} M A. Farhan, G. Lee and J. H. Shim, J. Phys.: Condens. Matter \textbf{26}, 042201 (2014).
      \bibitem{You2018} J. S. You, I. Lee, E. S. Choi, Y. J. Jo, J. H. Shim, J. S. Kim, Curr. Appl. Phys. In press, (2018).
   \bibitem{Rahn2017} M. C. Rahn, A. J. Princep, A. Piovano, J. Kulda, Y. F. Guo, Y. G. Shi, and A. T. Boothroyd,  Phys. Rev. B, \textbf{95}, 134405 (2017).
\bibitem{Wang2011} J. K. Wang, L. L. Zhao, Q. Yin, G. Kotliar, M. S. Kim, M. C.
Aronson, and E. Morosan, Phys. Rev. B \textbf{84}, 064428 (2011).
\bibitem{Ray2017} S. J. Ray, and L. Alff, Phys. Status Solidi B, \textbf{254}, 1600163 (2017).
   \bibitem{Huang2017} S. Huang, \textit{et al.}., PNAS, \textbf{114}, 6256 (2017).
 \bibitem{Liu2016} J. Liu, \textit{et al.}., Sci Rep., \textbf{6}, 30525 (2016).
\bibitem{Guo2014} Y. F. Guo, A. J. Princep, X. Zhang, P. Manuel, D. Khalyavin, I. I. Mazin, Y. G. Shi, A. T. Boothroyd, Phys. Rev. B, \textbf{90}, 075120 (2014).
  \bibitem{AZhang2015} A. Zhang, C. Liu, C. Yi, G. Zhao, T. Xia, J. Ji, Y. Shi, R. Yu, X. Wang, C. Chen and Q. Zhang, Nat. Commun. \textbf{7}, 13833 (2015).
  \bibitem{Toth2015} S. Toth and B., Lake, J. Phys.: Condens. Matter \textbf{27}, 166002 (2015).
 \bibitem{Blochl1994}P. E. Bl{\"o}chl, Phys. Rev. B {\bf 50}, 17953 (1994). 
\bibitem{Kresse1996}G. Kresse and J. Furthm{\"u}ller, Phys. Rev. B {\bf 54}, 11169 (1996). 
\bibitem{Kresse1999}G. Kresse and D. Joubert, Phys. Rev. B {\bf 59}, 1758 (1999). 
\bibitem{Taroni2008} A. Taroni, S. T. Bramwell and P. C. W. Holdsworth, J. Phys.: Condens. Matter \textbf{20}, 275233 (2008).
  \bibitem{Zhang2016} Q. Zhang,C. M. N. Kumar,  W. Tian,  K. W. Dennis,  A. I. Goldman,  and D. Vaknin, Phys. Rev. B, \textbf{93}, 094413 (2016).   
     \bibitem{Zhang2015} Q. Zhang, W. Tian, S. G. Peterson, K. W. Dennis, and D. Vaknin,  Phys. Rev. B, \textbf{91}, 064418 (2015).
 \bibitem{Satoshi2018} By mapping the total energy of different magnetic ordering to that of a Heisenberg-type model consisting of in-plane (out-of-plane) exchange $J_{1}$ between Mn moments $S=1.8 \mu_B$ with
 the single ion anisotropy $K$, we found $J_1= 25.0$ (AFM), $J_{\perp}=-0.23$ (FM), and $K = -0.16$ (uniaxial along the $a$) in unit of meV. The precise values of these parameters depend on the value of local $U$ on a Mn site and the exchange correlation potential, which could reconcile why these values are larger than those obtained from the fits to experimental spin waves. 
\bibitem{Young2015}S. M. Young and C. L. Kane, Phys. Rev. Lett. {\bf 115}, 126803 (2015). 


\end{thebibliography}
\end{document}